\documentstyle[epsfig,twocolumn,pra,aps]{revtex}
\begin{document}
\draft
\twocolumn[\hsize\textwidth\columnwidth\hsize\csname @twocolumnfalse\endcsname

\title{Transfer of nonclassical features in quantum teleportation \\
via a mixed quantum channel}

\author{Jinhyoung Lee,$^{1,2}$ M. S. Kim,$^{1}$ and Hyunseok Jeong$^2$}

\address{$^1$ School of Mathematics and Physics, The Queen's
  University of Belfast, BT7 1NN, \\ 
  United Kingdom \\ 
  $^2$ Department of Physics, Sogang University, CPO Box 1142, Seoul
  100-611, Korea}

\date{\today} 

\maketitle

\begin{abstract}  
  Quantum teleportation of a continuous-variable state is studied for
  the quantum channel of a two-mode squeezed vacuum influenced by a
  thermal environment.  Each mode of the squeezed vacuum is assumed to
  undergo the same thermal influence.  It is found that when the mixed
  two-mode squeezed vacuum for the quantum channel is separable, any
  nonclassical features, which may be imposed in an original unknown
  state, cannot be transferred to a receiving station.  A two-mode
  Gaussian state, one of which is a mixed two-mode squeezed vacuum, is
  separable if and only if a positive well-defined $P$ function can be
  assigned to it.  The fidelity of teleportation is considered in
  terms of the noise factor given by the imperfect channel.  It is
  found that quantum teleportation may give more noise than direct
  transmission of a field under the thermal environment, which is due
  to the fragile nature of quantum entanglement of the quantum
  channel.
\end{abstract}


\vskip1pc] 

\section{Introduction}
\label{sec:int}

Quantum teleportation is one of the important manifestations of
quantum mechanics.  By quantum teleportation an unknown quantum state
is destroyed at a sending station while its replica state appears at a
remote receiving station via dual quantum and classical channels.  The
key to quantum teleportation is the entanglement of the quantum
channel.  Quantum teleportation has been studied for various systems
including two-level systems \cite{Bennett93}, $N$-dimensional systems
\cite{Stenholm98}, and continuous variables
\cite{Vaidman94,Braunstein98,Ralph98}.  In particular, quantum
teleportation of continuous variable states has been at a focus
because of a high detection efficiency and handy manipulation of
continuous variable states \cite{Braunstein98,Furusawa98}.

Quantum teleportation of a continuous-variable state was first
suggested by Vaidman employing the Einstein-Podolsky-Rosen (EPR) state
\cite{EPR35} for the quantum channel in the framework of nonlocal
measurements \cite{Vaidman94}.  Braunstein and Kimble made a use of
quadrature-phase entanglement in a two-mode squeezed vacuum to
teleport the quadrature-phase variables.  With the high detection
efficiency of the homodyne measurement and highly squeezed light,
Ralph and Lam \cite{Ralph98} and Furusawa {\em et al}.
\cite{Furusawa98} realized quantum teleportation of continuous
variable states by experiments.  Ralph and Lam produced the required
entangled state using two bright squeezed sources.  A two-mode
squeezed vacuum is entangled with respect not only to quadrature
phases but also to photon-number difference and phase sum.  Based on
this number-phase entanglement, Milburn and Braunstein suggested
another protocol to teleport a continuous variable state
\cite{Milburn99}.

There are a few problems in the quantum teleportation of
quadrature-phase variables using the two-mode squeezed vacuum.  The
perfect quantum teleportation is possible only with a maximally
entangled state which means infinite squeezing in the squeezed state.
The mean energy of a two-mode squeezed state increases exponentially
as the squeezing increases so that the maximally entangled squeezed
state is unphysical.  As the quantum channel is exposed to the real
world, it is influenced by the environment, which turns the {\em pure}
squeezed state into a {\em mixture} and deteriorates the entanglement
property.  The environmental effect is unavoidable for any type of
teleportation and there have been suggestions to purify mixed
entangled state into a maximally entangled singlet state for a
discrete two-level system \cite{Bennett96}.  Duan {\it et al.} suggested a
way to purify a Gaussian continuous variable state \cite{Duan99-1}.
However, their purification protocol may concentrate entanglement only
to a finite dimensional Hilbert space.  In fact, it is impossible to
purify a two-mode squeezed state into a maximally entangled state as
it is unphysical.  Opartny {\em et al.} showed that the problem of not
having the maximally entangled squeezed vacuum can be overcome by
conditional measurements \cite{Opartny99}.  Entanglement
quantification and purification for continuous-variable states has
been studied by Parker {\em et al.} \cite{Parker00}. The imperfect
detection efficiency and the imperfect realization of unitary
transformation at the receiving station can also lower the efficiency
of teleportation.

In this paper, we are interested in the efficiency of quantum
teleportation in the real world.  Nonclassical properties such as
sub-Poissonicity and squeezing of the original state can be very
useful for communication purposes.  As the quantum channel is not
maximally entangled, some or all of the nonclassical properties can be
lost during the teleportation.  Braunstein and Kimble found that when
the quantum channel is not squeezed, i.e., when the channel is merely
a two-mode vacuum, no quantum features can be observed in the
teleported state \cite{Braunstein98}.  This is due to quantum tariffs
of vacuum noise, which arises at the sending and receiving stations.
The tariff was coined as {\em quduty} by Braunstein and Kimble.  The
pure two-mode squeezed state becomes mixed as the quantum channel is
embedded in the environment.  Quantum teleportation via the mixed
channel can bear a different nature.  For example, one may ask ``Does
the classical correlation play any role to transfer the nonclassical
features?''  It is not clear so far under which condition any
nonclassical features implicit in an original unknown state cannot be
transferred by teleportation via a mixed channel.  We also consider
the fidelity of teleportation to measure how
close the teleported state is to the original state when the quantum
channel is mixed.  Popescu studied quantum teleportation of a discrete
two-level system for a mixed quantum channel and found that even when
the quantum channel is not maximally entangled, it has the fidelity
better than any classical teleportation protocol \cite{Popescu94}.  In
this paper, we restrict ourselves to the situation that the
decoherence effect is the same on each mode of the two-mode squeezed
vacuum.

The continuous variable state can be easily analyzed using the
quasiprobability functions \cite{Hillery84}.  The description of a
quantum-mechanical state in phase space is not unique due to the
uncertainty principle; hence there are a family of quasiprobability
functions of which the $P$, $Q$, and Wigner functions are widely used
\cite{Cahill69}.  In particular, it is well-known that the $P$
function can be used as a measure of the nonclassicality of a given
field because the $P$ function is positive well-defined only for a
classical state \cite{Mandel86}.  The nonclassical depth is defined
based on how much noise to put into the nonclassical state to have a
positive well-defined $P$ function.

When teleportation is imperfect, a noise-added replica state is
produced at the receiving station.  By analyzing the added noise, we
find the critical point for the quantum channel not to transfer any
nonclassical features which may be imposed in an original unknown
state.  We examine the coincidence of the critical point with the
moment when the quantum channel becomes separable.  To do that we find
the necessary and sufficient condition of separability for any
two-mode Gaussian state \cite{Duan99}, one of which is the mixed
two-mode squeezed state.  The fidelity, which is defined as the inner
product of the original and teleported states, can be represented by
the overlap of their Wigner functions \cite{Aharonov80}.  We show that
the fidelity is a function of the added noise.

The added noise by teleportation is compared with that by direct
transmission of the original state.  It is found that the nonclassical
nature of the original state can be more easily lost by teleportation
than by direct transmission.  This is because teleportation relies on
the entanglement of the quantum channel, which is very fragile.

\section{Quasiprobability functions}

Before considering quantum teleportation, we briefly introduce the
quasiprobability functions.  The family of quasiprobability functions
are obtained by the following convolution relation
\begin{equation}
  \label{eq:gqf}
  R_\sigma (\alpha) = \int d^2\beta \left[
    \frac{2}{\pi(1-\sigma)}\exp \left( - \frac{2\left| \alpha-\beta
        \right|^2}{1-\sigma} \right) \right] P(\beta)
\end{equation}
where the $\sigma$-parameterized $R_\sigma(\alpha)$ function becomes
$Q$ function for $\sigma=-1$, Wigner ($W$) function for $\sigma=0$,
and $P$ function for $\sigma=1$.  By the Fourier transform, we find
the relation between their characteristic functions
\begin{equation}
C_{\sigma}^R(\xi) = \exp\left[-\frac{(1-\sigma)|\xi|^2}{2}\right]C^P(\xi)
\label{characteristic}
\end{equation}
where $C_{\sigma}^R (\xi)$ and $C_{\sigma}^P (\xi)$ are the
characteristic functions for the $R$ and $P$ functions, respectively.
The family of two-mode quasiprobability functions can be analogously
defined as
\begin{eqnarray}
\label{gdf2}
R_{\sigma}(\alpha,\beta) = & &\frac{4}{\pi^2(1-\sigma)^2}\int d^2 \phi
      d^2 \eta \nonumber\\
  & & \times  \exp \left( - \frac{2\left| \alpha-\phi
        \right|^2}{1-\sigma}- \frac{2\left| \beta-\eta
        \right|^2}{1-\sigma} \right) P(\phi,\eta).
\end{eqnarray}

\section{Teleportation for continuous variables in thermal environments}
\label{sec:cvt}

A continuous variable state $\hat{\rho}_o$ can be teleported with use
of a two-mode squeezed vacuum for a quantum channel
\cite{Braunstein98}.  Two modes $b$ and $c$ of the squeezed vacuum are
distributed separately to a sending and a receiving stations. The
protocol comprises two operations at the sending station and one
operation at the receiving station.  At the sending station, the
original unknown state of mode $a$ is mixed with a mode $b$ of the
quantum channel by a 50/50 beam splitter.  Two conjugate quadrature
variables are measured respectively for the two output fields of the
beam splitter.  The measurement results are sent to the receiving
station through the classical channel.  The other mode $c$ of the
squeezed vacuum is then displaced at the receiving station according
to the measurement results.  It is important to displace the photon of
mode $c$ entangled with the photon measured at the sending station.
Braunstein and Kimble considered the teleportation protocol for the
{\em pure} state of the quantum channel \cite{Braunstein98}.  In this
paper we investigate the teleportation via the mixed quantum channel
to consider the influence of a thermal environment.  We assume that
the thermal environment gives the same effect on each mode of the
quantum channel and the original state is prepared in a pure state.

The two-mode squeezed vacuum of the quantum channel is entangled and
represented by the Wigner function \cite{Barnett87}
\begin{eqnarray}
\label{eq:wig2mode}
W_{qc}(\alpha_b, \alpha_c)= & &\frac{4}{\pi^2} \exp\left[-2 \left(
    |\alpha_b|^2+|\alpha_c|^2 \right) \cosh 2s_{qc} \right. \nonumber \\
  & & \left. + 2\left( \alpha_b
    \alpha_c + \alpha_b^* \alpha_c^*\right) \sinh 2s_{qc} \right],
\end{eqnarray}
where $s_{qc}$ is the degree of squeezing and the complex quadrature
phase variable $\alpha_{b,c} = \alpha_{b,c}^r + i \alpha_{b,c}^i$.
When $s_{qc}\rightarrow\infty$ the state (\ref{eq:wig2mode}) manifests
the maximum entanglement and becomes an EPR state.  However, the mean
photon number, which is $2\sinh ^2 s_{sq}$, becomes infinity in this
limit.

Before the action of the beam splitter, the total state is a product
of the original state and the state of the quantum channel, which is
represented by the total Wigner function
$W_t(\alpha_a,\alpha_b,\alpha_c) = W_o(\alpha_a)
W_{qc}(\alpha_b,\alpha_c)$ where $W_o(\alpha_a)$ is the Wigner
function of the original state $\hat{\rho}_o$.  The product state of
the original field and quantum channel becomes entangled at the beam
splitter.  Considering the unitary action of the beam splitter, the
quadrature variables $\alpha_{d,e}$ of the output fields are related
to those of the input fields:
$\alpha_{d,e}=(\alpha_b\pm\alpha_a)/\sqrt{2}$.  The Wigner function
$W^B_t(\alpha_d,\alpha_e,\alpha_c)$ for the total field after the beam
splitter is
\begin{equation}
  \label{eq:rwfab}
  W_t^B\left(\alpha_d,\alpha_e,\alpha_c\right) = W_t \left(
    \frac{\alpha_e+\alpha_d}{\sqrt{2}},
    \frac{\alpha_e-\alpha_d}{\sqrt{2}}, \alpha_c\right) 
\end{equation}
which exhibits entanglement between the modes $a$ and $b$.

Setting homodyne detectors at the output ports of the beam splitter,
the imaginary part of $\alpha_d$ and the real part of $\alpha_e$ are
simultaneously measured by appropriately choosing the phases of
reference fields for the homodyne detectors. Each measurement result
is transmitted to the receiving station to displace the quadrature
variables of the field mode $c$.  We have to make it sure that the
displacement operation is done on the photon of mode $c$ entangled
with the photon measured at the sending station.  After the displacement
$\Delta(\alpha_d^i, \alpha_e^r)$ the field of mode $c$ becomes to be
represented by the Wigner function $W_r(\alpha_c)$;
\begin{equation}
  \label{eq:rwf}
  W_r(\alpha_c) = \int d^2\alpha_d~d^2\alpha_e
  W_t^B(\alpha_d,\alpha_e,\alpha_c-\Delta(\alpha_d^i, \alpha_e^r)). 
\end{equation}
Braunstein and Kimble \cite{Braunstein98} found that the displacement
of $\Delta(\alpha_d^i,\alpha_e^r)=-\sqrt{2}(\alpha_e^r-i\alpha_d^i)$
maximizes the fidelity when the channel is a pure two-mode squeezed
state.  The probability $P(\alpha_d^i,\alpha_e^r)$ of measuring
$\alpha_d^i$ and $\alpha_e^r$ at the sending station is the same as
the marginal Wigner function
\begin{eqnarray}
  P(\alpha_d^i, \alpha_e^r) &=& \int d\alpha_d^r~d\alpha_e^i
  d^2\alpha_c W_t^B(\alpha_d,\alpha_e,\alpha_c). 
\end{eqnarray}

\subsection{two-mode squeezed vacuum in thermal environments}

The quantum channel initially in the two-mode squeezed vacuum results
in a mixed state by the interaction with the thermal environment.
Assuming that two thermal modes are independently coupled with the
quantum channel the dynamics of the squeezed field is described by a
Fokker-Planck equation in the interaction picture \cite{Jeong99}
\begin{eqnarray}
  \label{eq:MasterEq}
  \frac{\partial W_{qc}(\alpha_b,\alpha_c;t)}{\partial t} = & & {\gamma
    \over 2} \sum_{i=b,c} \left [ \frac{\partial}{\partial \alpha_i}
    \alpha_i + \frac{\partial}{\partial \alpha_i^*} \alpha_i^* \right.
    \nonumber \\
    & & \left. +
    \left( 1+2\bar{n} \right) \frac{\partial^2}{\partial \alpha_i
      \partial \alpha_i^*}\right] W_{qc}(\alpha_b,\alpha_c;t),
\end{eqnarray}
where $\bar{n}$ is the average photon number of the thermal
environment.  The two thermal modes are assumed to have the same
average energy and coupled with the channel in the same strength.
This assumption is reasonable as the two modes of the squeezed state
are in the same frequency and the temperature of the environment is
normally the same.  By solving the Fokker-Planck equation
(\ref{eq:MasterEq}), the time-dependent Wigner function is obtained as
\cite{Jeong99}
\begin{eqnarray}
  \label{eq:Wignerweget}
  W_{qc}(\alpha_b,\alpha_c; T)={\cal N} \exp & & \left[
    -\frac{2\Gamma}{\Gamma^2-\Lambda^2} \left( |\alpha_b|^2 +
      |\alpha_c|^2 \right) \right. \nonumber \\
    & & \left. + \frac{2\Lambda}{\Gamma^2-\Lambda^2}
    \left( \alpha_b \alpha_c + \alpha_b^* \alpha_c^* \right) \right]
\end{eqnarray}
where ${\cal N}$ is the normalization factor and two parameters,
$\Gamma= T (1+2\bar{n})+(1-T)\cosh 2s_{qc}$, $\Lambda= (1-T)\sinh
2s_{qc}$.  The renormalized time $T(t) = 1-\exp(-\gamma t)$. The
relative strength of $\Lambda$ to $\Gamma$ determines how much the
mixed channel is entangled.  When $\Lambda$ is zero for $T \rightarrow
1$, the channel loses any correlation so to have neither classical nor
quantum correlation.  At $T=0$ the mixed squeezed state
(\ref{eq:Wignerweget}) is simply the squeezed vacuum
(\ref{eq:wig2mode}).

When the quantum channel is embedded in thermal environments, the
teleported state is still represented by the Wigner function
(\ref{eq:rwf}) with the quantum channel (\ref{eq:Wignerweget}).
However, a question remains in the unitary operation at the receiving
station when the channel is a mixed state.  According to the
philosophy of the faithful teleportation, the displacement has to be
determined to maximize the fidelity of teleportation.  The fidelity
${\cal F}$, which measures how close the teleported state is to the
original state, is the projection of the original pure state
$|\Psi_o\rangle$ onto the teleported state of the density operator
$\hat{\rho}_r$: ${\cal F}=\langle\Psi_o|\hat{\rho}_r|\Psi_o\rangle$.
The fidelity is also represented by the overlap between the Wigner
functions for the original and teleported states \cite{Aharonov80};
\begin{equation}
  \label{eq:fidelity}
  {\cal F} = \pi \int d^2 \alpha W_o(\alpha) W_r(\alpha).
\end{equation}
For a maximally entangled quantum channel, the original pure state is
reproduced at the receiving station and the fidelity is unity.  For an
impure or partially entangled channel, the unitary operation at the
receiving station may depend on original states to maximize the
fidelity, which has been shown for the teleportation of a two-level
state \cite{Popescu94,Lee99b}.  For the infinite dimensional Hilbert
space, a formal study is very much complicated.  However, we have
found that even when the channel is mixed, the displacement of
$\Delta(\alpha_d^i,\alpha_e^r)= - \sqrt{2}(\alpha_e^r-i\alpha_d^i)$
maximizes the fidelity for a coherent projector
$|\mu\rangle\langle\nu^*|$, where $|\mu\rangle$ and $|\nu^*\rangle$
are coherent-state bases.  An unknown state can be written as a
weighted integral of the coherent projection operators
\begin{equation}
\label{positive-p}
\hat{\rho}_o=\int d^2\mu d^2\nu P_o(\mu,\nu) |\mu\rangle\langle\nu^*|
\end{equation}
where $P(\mu,\nu)$ is proportional to the positive-$P$ function
\cite{Drummond80}. The unitary operation, which maximizes the
fidelity, at the receiving station is thus independent of the original
state.
\subsection{separability of the quantum channel}

A discrete bipartite system of modes $b$ and $c$ is separable when its
density operator is represented by $\hat{\rho}=\sum_r P_r
\hat{\rho}_{b,r}\otimes\hat{\rho}_{c,r}$.  Separability and measures
of entanglement for continuous variable states has been studied
\cite{Parker00,Duan99}.  In particular, Duan {\em et al.} found a
criterion to determine separability of a two-mode Gaussian state.
Here, we have a somewhat different approach to find when a two-mode
squeezed vacuum in thermal environments is separable and not
quantum-mechanically entangled.  Our analysis of separability for the
mixed squeezed vacuum is extended and fully described for any two-mode
Gaussian state in Appendix.

As shown in Appendix, the mixed two-mode squeezed vacuum in the
thermal environment is separable when a positive definite $P$ function
can be assigned to it.  The mixed two-mode squeezed vacuum serving the
quantum channel can then be written by a statistical mixture of the
direct-product states;
\begin{equation}
\label{cont-sep}
  \hat{\rho}_{qc}=\int d^2\beta {\cal P}(\beta) \hat{\rho}_{b}(\beta)\otimes
\hat{\rho}_{c}(\beta)
\end{equation}
where ${\cal P(\beta)}$ is a probability density function.

With use of Eqs.~(\ref{gdf2}) and (\ref{eq:Wignerweget}), we find that
the mixed two-mode squeezed vacuum is separable when $n_\tau=1$ where
$n_\tau$ is defined as
\begin{eqnarray}
  \label{cnot}
  n_\tau(\bar{n},s_{qc},T) &\equiv& \Gamma -
  \Lambda \nonumber\\ &=& 
  \left(2\bar{n} + 1 \right) T + (1 - T) \exp(-2s_{qc}) 
\end{eqnarray}
according to the condition (\ref{eq:cfpf}).  This is in agreement with
Duan {\em et al.}'s separation criterion \cite{Duan99}.  The pure
two-mode squeezed vacuum for $T=0$, is never separable unless
$s_{qc}=0$.  For the zero temperature environment, i.e., $\bar n =0$,
the two-mode squeezed state stays quantum-mechanically entangled at
any time.  For the reasons given in Sec. IV, we call $n_\tau$ as the
noise factor.

If $n_{\tau} <1$, the quantum channel state is entangled and the
teleportation is performed at the quantum level.  When $n_{\tau} \ge
1$, the quantum channel is no longer quantum-mechanically entangled.
However the inter-mode correlation is still there as $\Lambda \neq 0$
in Eq.~(\ref{eq:Wignerweget}).  Questions then arise: Does this
classical correlation influence the teleportation?  Can any
nonclassical properties imposed in an original state be teleported by
the classically-correlated channel?  Braunstein and Kimble found that
when a pure two-mode squeezed state is separable, i.e., $s_{sq}=0$,
observation of any nonclassical features in the teleported state is
precluded.  However, when a pure state is separable there is no
classical correlation either.

\section{Transfer of Nonclassical Features}
\label{sec:tnf}

An imperfect replica state is reproduced at the receiving station when
the quantum channel is not maximally entangled.  It is well known that
any linear noise-addition process, for example linear dissipation and
amplification, can be described by the convolution relation of the
quasiprobability functions\cite{MSKim95}.  With use of the Wigner
functions for an arbitrary original state (\ref{positive-p}) and for
an impure quantum channel (\ref{eq:Wignerweget}), we find that the
equation (\ref{eq:rwf}) leads to the following convolution relation
\begin{equation}
  \label{eq:cltw}
  W_r(\alpha) = \int d^2\beta P_\tau(\alpha-\beta) W_o(\beta)
\end{equation}
where the function $P_\tau$ characterizes the teleportation process;
\begin{equation}
  \label{eq:pct}
  P_\tau(\alpha-\beta) = \frac{1}{\pi n_\tau} \exp{\left( -
      \frac{1}{n_\tau} \left|\alpha-\beta\right|^2 \right)}
\end{equation}
and the noise factor $n_\tau$, defined in Eq.~(\ref{cnot}), is
completely determined by the characteristics of the quantum channel.
The noise factor increases monotonously as the interaction time $T$
increases.  The larger the initial squeezing, the less vulnerable the
quantum channel is.

The noise factor $n_\tau$ is related to the fidelity.  With use of
Eqs.~(\ref{eq:fidelity}) and (\ref{eq:cltw}) the fidelity can be
written as
\begin{equation}
  \label{eq:rnnf}
  {\cal F} = \pi \int d^2\alpha d^2\beta W_o(\alpha) P_\tau(\alpha-\beta)
  W_o(\beta).
\end{equation}
In the limit of $n_\tau \rightarrow 0$, the function
$P_\tau(\alpha-\beta)$ in Eq.~(\ref{eq:pct}) becomes a delta function
and the fidelity becomes unity.  The teleportation loses the original
information completely with ${\cal F}=0$ in the limit of
$n_\tau\rightarrow\infty$.

The properties of the nonclassical states have been calculated and
illustrated by quasiprobability functions. The nonclassical features
are associated especially with negative values and singularity of the
quasiprobability $P$ function \cite{Mandel86,Lutkenhaus95,Janszky96}.
Suppose an original state whose $P$ function is not positive
everywhere in phase space.  When this state is teleported, its
nonclassical features are certainly transferred to the teleported
state if the teleportation is perfect. If the teleportation is poor,
the teleported state may have its $P$ function positive definite and
lose the nonclassical features.

By the Fourier transform of Eq.~(\ref{eq:cltw}), the convolution
relation is represented in terms of the characteristic functions as
\begin{equation}
\label{character-noncl}
C_r^W(\xi) = \exp(-\bar n_\tau |\xi|^2) C_o^W(\xi).
\end{equation}
Using the relation (\ref{characteristic}) between characteristic
functions, Eq.(\ref{character-noncl}) is written as
\begin{equation}
\label{character-noncl2}
C_r^P(\xi) =\exp[-(n_\tau - 1) |\xi|^2] C_o^Q(\xi),
\end{equation}
where $C_o^Q(\xi)$ is the characteristic function for
$R_{\sigma=-1}(\alpha)$ of the original state.  The $P$ function is
not semi-positive definite if its characteristic function $C_r^P(\xi)$
is not inverse-Fourier-transformable.  Even when it is
inverse-Fourier-transformable, there is a chance for the $P$ function
to become negative at some points of phase space.  L\"{u}tkenhaus and
Barnett found that only when $\sigma \le -1$ the quasiprobability
$R_\sigma(\alpha)$ for any state is semi-positive definite.  We are
sure that, for any original state, the left-hand side of
Eq.~(\ref{character-noncl2}) is inverse-Fourier transformed to a $P$
function semi-positive definite only when $n_\tau \ge 1$.  This
condition is the same as the separability condition (\ref{cnot}) for
the quantum channel.  We conclude that {\em when a quantum channel is
  separable, i.e., not quantum-mechanically entangled, no nonclassical
  features implicit in an original state is transferred by
  teleportation.} In other words, nonclassical features are not
teleported via a classically-correlated channel.

There are two well-known nonclassical properties which a
continuous-variable state may have: Sub-Poissonian photon statistics
and quadrature squeezing.  The two nonclassical properties have been
studied for noiseless communication.  We analyze the transfer of these
properties by teleportation in the following subsections.

\subsection{sub-Poissonian statistics and Fock state}

A state is defined to be sub-Poissonian when its photon-number
variance $(\Delta N)^2$ is smaller than its mean photon number $\bar
N$.  The expectation value of an observable for a state can be
obtained by use of the characteristic function $C^P (\xi)$ for its $P$
function \cite{Cahill69};
\begin{equation}
\label{observable}
\langle (\hat{a}^\dagger)^m \hat{a}^n \rangle
      =\left. \frac{\partial^m}{\partial\xi^m} 
      \frac{\partial^n}{\partial
      (-\xi^*)^n}C^P(\xi)\right|_{\xi=\xi^*=0}. 
\end{equation}
Substituting Eq.~(\ref{character-noncl2}) into Eq.~(\ref{observable}),
we find that the teleported state is sub-Poissonian when the noise
factor,
\begin{equation}
  \label{eq:rslp}
  n_\tau < \sqrt{\bar N_o^2+\bar N_o -(\Delta N_o)^2} - \bar{N}_o.
\end{equation}
where $\bar{N}_o$ and $(\Delta N_o)^2$ are the mean photon number and
photon-number variance for the original state.  If the original state
is Poissonian or super-Poissonian, the right-hand side of the
inequality is either negative or imaginary so the teleported state is
never sub-Poissonian.

Assuming the largest sub-Poissonicity, $(\Delta N_o)^2=0$, for the
original state, it is found that when the noise factor $n_\tau<
\sqrt{\bar{N}_o^2+\bar{N}_o}-\bar{N}\le 1/2$, some sub-Poissonian
property is found in the teleported state.  Thus, if the noise factor
of the quantum channel is larger than or equal to 1/2, the transfer of
any sub-Poissonian property is precluded.

A Fock state $|m\rangle$ has a definite energy and its photon-number
variance is zero.  When this extreme sub-Poissonian field is
teleported, the mean photon number and mean variance are
$\bar{N}_r=m+n_\tau$ and $\Delta N_r^2 = (2m+1)n_\tau+n_\tau^2$ at the
receiving station.  The Fock state $|m\rangle$ is written in the
Wigner representation as
\begin{eqnarray}
  \label{eq:exfsi}
  W_o(\alpha,m) = \frac{2}{\pi} (-1)^m \exp \left(-2|\alpha|^2\right)
  L_m\left(4|\alpha|^2\right).
\end{eqnarray}
where $L_m$ is a Laguerre polynomial.  From the convolution relation
(\ref{eq:cltw}), the teleported state is obtained as
\begin{eqnarray}
  \label{eq:exfs}
  W_r(\alpha) = & & \frac{2}{\pi}
  \frac{(2n_\tau-1)^{m}}{(2n_\tau+1)^{m+1}} \exp
  \left(-\frac{2|\alpha|^2}{2n_\tau+1}\right) \nonumber \\
  & & \times L_m\left(-\frac{4|\alpha|^2}{(2n_\tau)^2-1}\right).
\end{eqnarray}
The fidelity for the Fock state is given by Eq.~(\ref{eq:rnnf});
\begin{eqnarray}
  \label{eq:exfsf}
  {\cal F}_m = \frac{(1-n_\tau)^m}{(1+n_\tau)^{m+1}}
  P_m\left(\frac{1+n_\tau^2}{1-n_\tau^2}\right)
\end{eqnarray}
where $P_m$ is a Legendre polynomial. When $n_\tau=0$, ${\cal F}_m=1$.
In the limit of $n_\tau = 1$, where the teleportation is classical,
the fidelity ${\cal F}_m=(1/4)^{m}$ for $m \ne 0$. The vacuum state
has the fidelity ${\cal F}_0=1/2$ in the limit.

\subsection{quadrature squeezing and squeezed state}
 
We examine the transfer of quadrature squeezing which an unknown
original state may have.  The quadrature-phase operator is defined as
\begin{equation}
  \label{eq:qo}
  \hat{X}(\phi) = e^{-i\phi} \hat{a} + e^{i\phi} \hat{a}^{\dagger}
\end{equation}
where $\hat{a}$ ($\hat{a}^{\dagger}$) is an annihilation (creation)
operator and $\phi$ related to the angle in phase space.  A state is
said to be squeezed if the quadrature-phase variance $[\Delta
X(\phi)]^2 < 1$ for an angle $\phi$.  Substituting
Eq.~(\ref{character-noncl2}) into Eq.~(\ref{observable}), the mean
quadrature phase $\bar{X}(\phi)$ and variance $[\Delta X(\phi)]^2$ can
be calculated
\begin{equation}
  \label{eq:vos}
  \bar{X}_r(\phi) = \bar{X}_o(\phi);~~
  [\Delta X_r(\phi)]^2 = [\Delta X_o(\phi)]^2 + 2 n_\tau,
\end{equation}
where $\bar{X}_o(\phi)$ and $[\Delta X_o(\phi)]^2$ are the mean
quadrature phase and variance for the original state.  It is
interesting to realize that {\em the mean quadrature phase does not
  change at all during teleportation}.  This property holds regardless
of the channel entanglement.

The teleported state exhibits quadrature squeezing if
\begin{equation}
  \label{eq:ls}
  n_\tau < \frac{1}{2} \left\{ 1 - [\Delta X^2_u(\phi)]^2 \right\} \le {1\over 2}.
\end{equation}
Suppose that the original state has the absolute minimum variance
$[\Delta X_o (\phi')]^2=0$ at $\phi=\phi'$.  Then its teleported state
is also squeezed if the quantum channel is entangled enough to be
represented by the noise factor $n_\tau < 1/2$.  We note that the
condition $n_\tau < 1/2$ applies to the survival of both quadrature
squeezing and sub-Poissonian statistics.

A squeezed vacuum with the degree of squeezing $s_o$ is written in the
Wigner representation as
\begin{equation}
  \label{eq:exssi}
  W_o(\alpha) = \frac{2}{\pi} \exp \left[-2\exp(2s_o) \alpha_r^2 -
    2\exp(-2s_o) \alpha_i^2\right]
\end{equation}
where $\alpha_r$ and $\alpha_i$ are real and imaginary parts of
$\alpha$. Its teleported state is represented by the Wigner function
\begin{equation}
  \label{eq:exsst}
  W_r(\alpha) = \frac{2}{\pi\sqrt{A(s_o)A(-s_o)}} \exp
  \left[-\frac{2}{A(s_o)}\alpha_r^2 - \frac{2}{A(-s_o)}\alpha_i^2\right]
\end{equation}
where the parameter $A(s_o) = 2n_\tau + \exp(-2s_o)$. The fidelity is
given by
\begin{equation}
  \label{fous}
  {\cal F}(s_o) = \left(n_\tau^2 + 2n_\tau \cosh 2s_o +
    1\right)^{-1/2}.
\end{equation}
When the teleportation is classical with $n_\tau=1$, ${\cal F}(s_o) =
(2+2\cosh 2s_o)^{-1/2}$.

\section{Remarks}
\label{sec:rem}

Quantum teleportation can be made more reliable by sophisticated
schemes such as purification of the impure or partially entangled
quantum channel \cite{Parker00,Duan99-1}, detection with perfect
efficiency, and well-defined unitary operation.  However, in the real
world, the influence of noise cannot easily be disregarded.  We have
been interested in the influence of noise on the transfer of
nonclassicalities which may be imposed in an original unknown state.
To make the problem simple while honoring the real experimental
situation, we assumed that the same amount of noise affects the two
modes of the quantum channel.  We found that when the quantum channel
is separable the transfer of any nonclassicality is impossible:
Nonclassical features can not be teleported via a
classically-correlated channel.  The separability of a two-mode
Gaussian state is considered using the possibility of assigning a
positive well-defined $P$ function to the state after some local
unitary operations. We have analyzed the transfer of well-known
nonclassical features such as sub-Poissonicity and quadrature
squeezing.  The teleportation of the two nonclassical features is
ruled out under the same noise level.  The faithfulness of the
teleportation has also been discussed and the fidelities have been
found for the initial Fock state and squeezed state.  Because one of
the important ingredients of teleportation is that the original state
is {\em unknown} at the sending station.  Thus our measure of noise
factor $n_\tau$, which depends only on the quality of the channel, is
important.  Of course, to represent the quality of the teleportation
by a fidelity we have to know the average fidelity for the
teleportation, which is under investigation.

One question still arises: Is the teleportation better than the direct
transmission to transfer a nonclassical field?  A field may be
deteriorated by the thermal environment during the direct
transmission.  Solving a similar Fokker-Planck equation to
Eq.~(\ref{eq:MasterEq}) for a {\em single-mode} field, we find that,
by the direct transmission, the Wigner function at the receiving
station can be represented by the same equation as Eq.~(\ref{eq:cltw})
with the different noise factor $n_d$ \cite{MSKim95}:
\begin{equation}
\label{n-d}
n_d=\bar n T.
\end{equation}
Assuming that the imperfect teleportation is caused only by the impure
quantum channel embedded in the thermal environment, we compare the
two noise factors $n_\tau$ in Eq.~(\ref{cnot}) and $n_d$ in
Eq.~(\ref{n-d}).  We have implicitly assumed in this paper that the
two-mode squeezed state (quantum channel) generator is located in the
middle point between the sending and receiving stations.  The squeezed
photons in the quantum channel, thus, travel only a half distance
between the sending and receiving stations.  Bearing it in mind, we
find that
\begin{eqnarray}
\label{comparison}
n_\tau~(& &\mbox{for time}~t/2) - n_d~(\mbox{for time}~t) = \nonumber \\
& & \bar n[1-\sqrt{1-T}]^2 +1-\sqrt{1-T}[1-\exp(-2s_{qc})].
\end{eqnarray}
The right-hand side is semi-positive so that the noise given by
teleportation is more than that by direct transmission.  If we
consider that this result is obtained for the case when the other
operations including detection and unitary transformation in the
teleportation protocol is perfect, we conclude that {\em the
  nonclassical field is more robust in direct transmission than in
  teleportation}.  The reason is that the teleportation relies on
quantum entanglement of the quantum channel.  The quantum entanglement
based on inter-mode coherence is much more fragile than the
single-mode coherence.  However, the quantum teleportation can be
made more faithful by purification of the quantum channel while the
direct transmission does not have that possibility.

\acknowledgements 

This work was supported in part by the Brain Korea 21 grant (D-0055)
of the Korean Ministry of Education.

\appendix
\section*{positivity of $P$  function and separability for a
  Gaussian state}
\label{sec:aa}

A two-mode Gaussian state $\hat{\rho}$ of mode $b$ and $c$ is
separable when it is represented by a statistical mixture of the
direct-product states;
\begin{equation}
  \label{eq:tsss}
  \hat{\rho} = \int d^2\beta {\cal P}(\beta) \hat{\rho}_b(\beta)
  \otimes \hat{\rho}_c(\beta)
\end{equation}
where $\hat{\rho}_{b,c}(\beta)$ are density matrices, respectively,
for modes $b$ and $c$, and ${\cal P}(\beta)$ is a probability density
function with ${\cal P}(\beta) \ge 0$.  The states of
$\hat{\rho}_b(\beta)$ and $\hat{\rho}_c(\beta)$ can be nonclassical
and do not have to have their $P$ functions positive well-defined.
However, because they are Gaussian, it is possible to transform them
to assign positive well-defined $P$ functions by local unitary
transformations \cite{note1}.  The separable condition,
(\ref{eq:tsss}), can then be written as
\begin{eqnarray}
\label{condition-appendix1}
 \hat{\rho}^\prime = & & \int d^2\alpha_b \int d^2 \alpha_c \int
 d^2\beta {\cal P}(\beta) P(\alpha_b;\beta) 
 P(\alpha_c;\beta) \nonumber \\
 & & \times |\alpha_b\rangle \langle \alpha_b| \otimes |\alpha_c\rangle
 \langle \alpha_c| 
\end{eqnarray}
where $P_b(\alpha_b;\beta)$ and $P_c(\alpha_c;\beta)$ are the $P$
functions, respectively, for the fields of modes $b$ and $c$ after
some local unitary operations.  $\hat{\rho}^\prime$ is for the
two-mode Gaussian state after the local operations.

We want to prove in this appendix that if and only if when a two-mode
Gaussian state is separable, a positive well-defined $P$ function
$P(\alpha_b,\alpha_c)$ is assigned to it after some local unitary
transformations.

consider the sufficient condition. If a two-mode Gaussian state
$\hat{\rho}$ is separable, it can be written as
Eq.~(\ref{condition-appendix1}) after some local operations.  Both
$P_{b}(\alpha_b;\beta)$ and $P_c(\alpha_c; \beta)$ are positive
well-defined and ${\cal P}(\beta)$ is a probability density function
so
\begin{equation}
\label{int-appendix}
\int d^2\beta{\cal P}(\beta) P(\alpha_b;\beta)
 P(\alpha_c;\beta)
\end{equation}
is a normalized positive function, which we can take as the positive
well-defined $P$ function $P(\alpha_b, \alpha_c)$.  We have proved
that if a two-mode Gaussian state is separable, it has a positive
well-defined $P$ function after some local unitary operations.

Now let us prove the necessary condition.  If the locally-transformed
two-mode Gaussian state is represented by a positive well-defined $P$
function $P(\alpha_b,\alpha_c)$, the separable condition
(\ref{condition-appendix1}) becomes
\begin{equation}
  \label{eq:prsc}
  P(\alpha_b,\alpha_c) = \int d^2\beta {\cal P}(\beta)
  P_b(\alpha_b;\beta) P_c(\alpha_c;\beta). 
\end{equation}
Further by some additional squeezing and rotation it is always
possible to have the rotationally-symmetric variance $[\Delta
\alpha_i(\phi)]^2$ for any angle $\phi$. After these transformations,
the positive well-defined $P$ function $P(\alpha_b,\alpha_c)$ can be
written as
\begin{eqnarray}
  \label{eq:ppgs}
  P(\alpha_b,\alpha_c) = & & {\cal N} \exp \Big[-
    \sum_{i,j=b,c} \alpha_i N_{ij} \alpha_j^* \nonumber \\
   & & + \sum_{i=b,c} \left( \alpha_i
      \lambda_i^* + \alpha_i^* \lambda_i \right) \Big]
\end{eqnarray}
where ${\cal N}$ is the normalization constant, $N_{ij}$ a Hermitian
matrix, and $\lambda_i$ a complex number. The linear terms of
$\alpha_i$ are not considered because they do not affect the proof.
In fact, they can always be removed by some local displacement
operations.  Eq.~(\ref{eq:ppgs}) can be written as
\begin{equation}
  \label{eq:ppgs2}
  P(\alpha_b,\alpha_c) = \frac{{\rm Det}N_{ij}}{\pi^2} \exp \left(-
    \sum_{i,j=b,c} \alpha_i N_{ij} \alpha_j^* \right)
\end{equation}
where ${\rm Det}N_{ij}$ is the determinant of the Hermitian matrix
$N_{ij}$. To find an expression in the form of Eq.~(\ref{eq:prsc}),
let us introduce an auxiliary field ($\beta$, $\beta^*$) enabling the
function $P(\alpha_b,\alpha_c)$ to be represented by a Gaussian
integral;
\begin{eqnarray}
  \label{eq:pfia}
  P(\alpha_b,\alpha_c) = & & \frac{{\rm Det}N_{ij}}{\pi^3} \int
  d^2\beta \exp \Big(- |\beta|^2 - E_b(\alpha_b,\beta) \nonumber \\
    & & - E_c(\alpha_c,\beta)\Big)
\end{eqnarray}
where
\begin{eqnarray}
  \label{eq:exp}
  E_b(\alpha_b,\beta) &=& \left(N_{bb} + |N_{bc}|^2\right)
  |\alpha_b|^2 - \alpha_b N_{bc}\beta^* \nonumber \\
 & & - \alpha_b^*N_{bc}^*\beta \\ 
  E_c(\alpha_c,\beta) &=& \left(N_{cc} + 1\right) |\alpha_c|^2 +
  \alpha_c\beta^* + \alpha_c^*\beta
\end{eqnarray}
The integrand in Eq.~(\ref{eq:pfia}) can now be decomposed into three
Gaussian functions each of which satisfies the normalization condition
because
\begin{equation}
  \label{eq:cfpf}
  N_{ii} > 0~~~{\rm and}~~~{\rm Det} N_{ij} > 0
\end{equation}
for positive well-defined $P(\alpha_b, \alpha_c)$ in
Eq.~(\ref{eq:ppgs}).  Taking
\begin{eqnarray}
  \label{eq:tfwo}
  P_b(\alpha_b; \beta) &=& \frac{M_b}{\pi} \exp \bigg( 
  -M_b|\alpha_b|^2 +\alpha_b N_{bc}\beta^*  \nonumber \\ 
  && +\alpha_b^* N_{bc}^*\beta  -\frac{|N_{bc}|^2}{M_b}
    |\beta|^2 \bigg)\\ 
  P_c(\alpha_c; \beta) &=& \frac{M_c}{\pi} \exp \bigg( -M_c|\alpha_c|^2
    -\alpha_c\beta^* -\alpha_c^*\beta \nonumber \\
 && -\frac{1}{M_c}|\beta|^2 \bigg)\\ 
  {\cal P}(\beta) &=& \frac{M_s}{\pi} \exp \left(-M_s|\beta|^2 \right)
\end{eqnarray}
where $M_b=N_{bb}+|N_{bc}|^2$, $M_c=N_{cc}+1$, and $M_s = {\rm
  Det}N_{ij}/(M_b M_c)$, the $P$ function is finally obtained in the
form of Eq.~(\ref{eq:prsc}).  It is clear that ${\cal P}(\beta)$ is
the positive probability density function and the two-mode Gaussian
state is separable if it can be transformed to have a positive
well-defined $P$ function by some local unitary operations.

\end{document}